\documentclass[%
 reprint,
 superscriptaddress,
 amsmath,amssymb,aip,
]{revtex4-2}

\usepackage{graphicx}
\usepackage{dcolumn}
\usepackage{bm}
\usepackage{xcolor}
\usepackage[utf8]{inputenc}
\usepackage[T1]{fontenc}
\usepackage{mathptmx}
\usepackage{siunitx}
\usepackage{silence}
\DeclareSIUnit\torr{Torr}

\begin{document}

\preprint{AIP/123-QED}

\title{Vacuum Characterization of a Compact Room-temperature Trapped Ion System}

\author{Yuhi Aikyo}
\thanks{Author to whom correspondence should be addressed: yuhi.aikyo@duke.edu}
\author{Geert Vrijsen}
\affiliation{Department of Electrical and Computer Engineering, Duke University, Durham, NC 27708, USA}
\author{Thomas W. Noel}
\affiliation{ColdQuanta, Inc., Boulder, CO 80301, USA}
\author{Alexander Kato}
\affiliation{Department of Physics, University of Washington, Seattle, WA 98105, USA}
\author{Megan K. Ivory}
\affiliation{Sandia National Laboratories, Albuquerque, New Mexico 87185, USA}
\author{Jungsang Kim$^{1,}$}
\affiliation{IonQ, Inc., College Park, MD 20740, USA}

\date{\today}%

\begin{abstract}
We present the design and vacuum performance of a compact room-temperature trapped ion system for quantum computing, consisting of a ultra-high vacuum (UHV) package, a micro-fabricated surface trap and a small form-factor ion pump. The system is designed to maximize mechanical stability and robustness by minimizing the system size and weight. The internal volume of the UHV package is only $\approx \SI{2}{\cubic\centi\meter}$, a significant reduction in comparison with conventional vacuum chambers used in trapped ion experiments. We demonstrate trapping of $^{174}$Yb$^+$ ions in this system and characterize the vacuum level in the UHV package by monitoring both the rates of ion hopping in a double-well potential and ion chain reordering events. The calculated pressure in this vacuum package is $\approx \SI{2.2e-11}{\torr}$, which is sufficient for the majority of current trapped ion experiments.
\end{abstract}

\maketitle

Systems using atomic ions are among the leading physical platforms for a practical quantum computer because of their long coherence times~\cite{Langer2005,wang2017}, full connectivity between qubits~\cite{Linke2017, murali_full-stack_2019, wright2019} and high-fidelity gate operations~\cite{Wineland1998,Monroe1164,wright2019}. However, unlike qubits based on solid state devices, the integration approach for scaling trapped ion systems is not obvious.  Many ideas for engineering complex trapped ion systems have been outlined to build practically useful trapped ion quantum computing systems~\cite{Monroe1164, brown_co-designing_2016, Lekitsche2017,Schwindt2016}.

Trapped ion experiments, whether they use traditional linear Paul Traps~\cite{Raizen1992} or micro-fabricated surface traps~\cite{PhysRevLett.96.253003}, ultimately rely on a lack of collision events with background gas molecules in order to perform reliable high-fidelity gates.  Critically, pressures in the ultra-high vacuum (UHV) regime ($\approx\SI{1e-11}{\torr}$) are required to keep the background gas collision rates low enough to minimize ion chain reordering events and loss of ions from the trap~\cite{brown_co-designing_2016}.  Additionally, quantum computation requires high fidelity gates, necessitating excellent opto-mechanical robustness and stability of a scalable trapped ion quantum computer. Optical frequencies of lasers driving near-resonant processes should be stabilized to a part in $10^{10}$ range in order to properly utilize these transitions for qubit manipulation and read-out~\cite{Olmschenk2007}.  Quantum logic gates are often driven using Raman transition, where two far-detuned non-co-propagating beams with precise frequency difference intersect at the location of the ion. Beam path length and pointing fluctuations of these Raman beams lead to optical phase and intensity fluctuations at the ions, which results in imperfect gates.  In order to avoid these problems, the trapped ions system and the optical elements used for the delivery of the laser beams should be stable against environment noise such as temperature fluctuations, air currents and mechanical vibrations. To address these requirements for a scalable trapped ion quantum computer, cryogenic systems were proposed and investigated~\cite{brown_co-designing_2016,pagano_cryogenic_2018,Antohi2009}, where the volume of the ultra-high vacuum (UHV) operating area can be minimized by taking advantage of cryogenic temperature while maintaining its vacuum quality. However, there are drawbacks of using cryogenic system, such as the size and cost of a cryogenic system and the vibrations generated by closed cycle cryostats.

In this paper we present a compact trapped ion system operating at room temperature. The system volume is reduced by several orders of magnitude below comparable ion trap systems, while the pressure is maintained in the UHV regime despite the high surface-area-to-volume ratio, making it a suitable system to extend to cryo environments. We present the detailed design of this compact room temperature system, and characterize the vacuum level of the system by measuring the well-to-well hopping rates of trapped $^{174}$Yb$^+$ ions in a double-well potential, as well as the occurrence rate of chain re-ordering collisions. The method and results of this vacuum characterization are presented in the end of this article.

\begin{figure*}
        \centering
        \includegraphics[width=150mm]{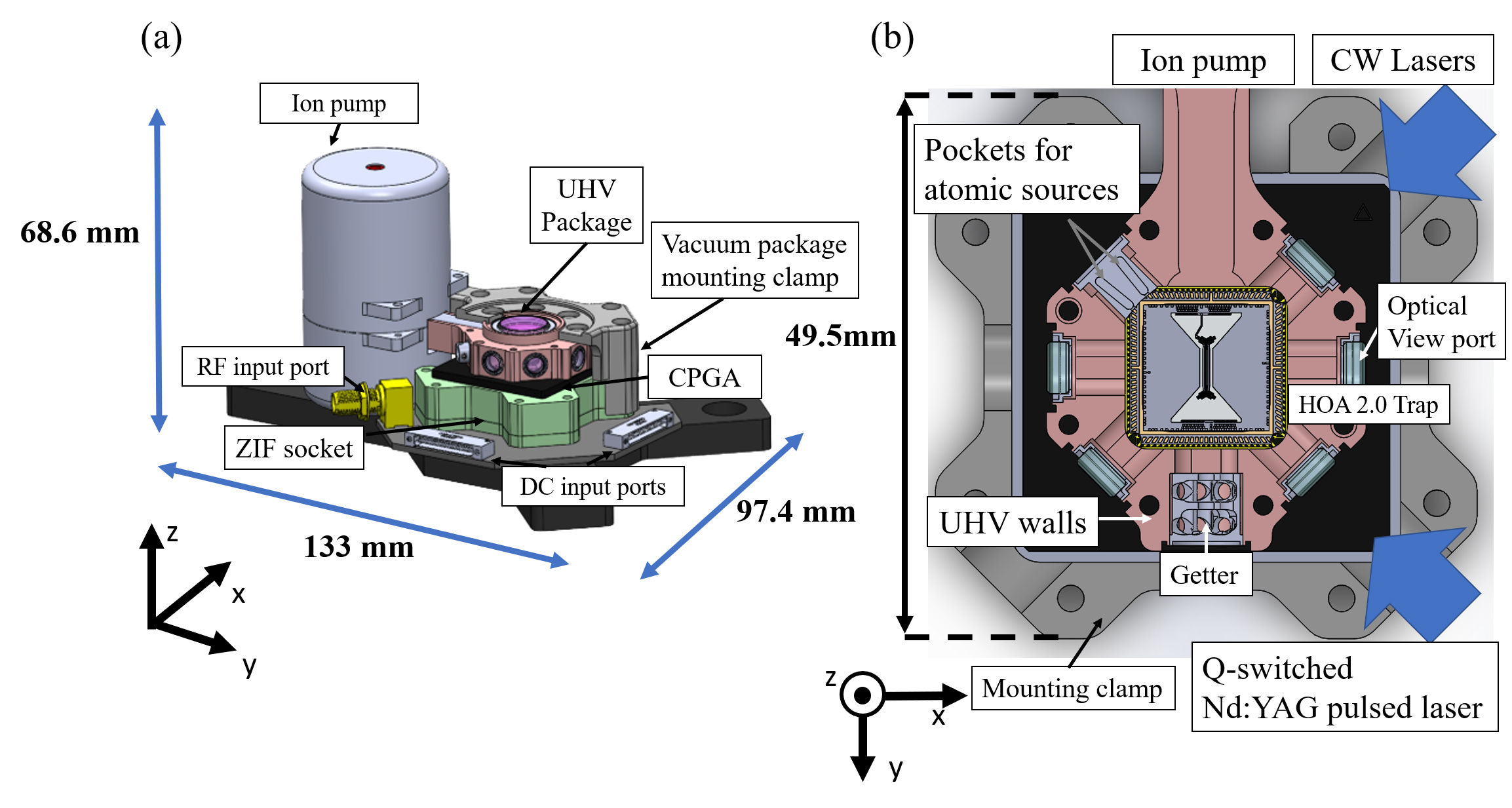}
        \caption[VISIT system]{(a) A 3-dimensional computer-aided design (CAD) model of the compact room temperature trapped ion system. The main system components are labeled. (b) A cross-sectional top view of the UHV package. The UHV walls (colored pink) are located inside the vacuum package mounting clamp (colored dark grey). The Yb atomic sources are held in the pockets located at the top left side of the UHV chamber.  The ablation laser is incident from the opposite side of the Yb source, and crosses the trap diagonal to the trap axis. The micro-fabricated surface trap (Sandia HOA~2.0) is mounted in the center of the UHV package, with the trap axis co-linear with the ion pump direction. The ions are trapped in the central region of the trap. The continuous wave (CW) lasers are applied from the top right side of the view-port to ions for photo-ionization (\SI{391}{\nano\meter}, \SI{399}{\nano\meter}), Doppler cooling and detection (\SI{370}{\nano\meter}), and optical re-pumping from long lived states (\SI{638}{\nano\meter}, \SI{935}{\nano\meter}).}
        \label{fig:visit}
\end{figure*}

The compact vacuum package is described schematically in Fig.~\ref{fig:visit}(a). A significant reduction in system volume is enabled by two improvements: 1) The ceramic pin grid array (CPGA, NKT CPG10039) which typically acts as an in-vacuum break-out board now is Au-Sn soldered to the ring flame of the package and acts as the electrical feedthrough between UHV and atmosphere; 2) Conflat flanges which typically provide viewports and electrical feedthroughs for additional components have been replaced with laser welded components with significantly reduced size. The vacuum system consists of the UHV package, the atomic source assembly, viewports, the CPGA, the ion pump, and the direct current (DC) and radio frequency (RF) feedthrough connectors, with an overall dimension of $\approx\SI{135}{\milli\meter}\times\SI{100}{\milli\meter}\times\SI{70}{\milli\meter}$. The UHV volume inside the package is $\approx\SI{2}{\cubic\centi\meter}$. The viewports, non-evaporable getter plug, and atomic source assembly are all welded to the vacuum housing using identical form-factor seals, making them interchangeable for flexibility during system design. Once the trap is die-attached and wire bonded in the CPGA, it is loaded to a UHV packaging facility. The UHV package (ColdQuanta P/N: CQARO-0100.1) is assembled with viewports, non-evaporable getter plug, atomic source, and custom-developed ion pump, and welded for hermeticity. The CPGA ringframe is bonded to the UHV housing by ColdQuanta using a proprietary bonding and alignment procedure. The UHV packaging station operates at $\approx \SI{1e-9}{\torr}$ vacuum levels, and supports self-aligned assembly with tolerances well within the requirements needed for the optical alignment of the beams in the experiment (~\SI{25}{\micro\meter}). Once sealed, the UHV package is removed from the assembly chamber and the ion pump is turned on to maintain the internal vacuum level to $\approx \SI{2e-11}{\torr}$.

\begin{figure}[ht]
        \centering
        \includegraphics[width=\columnwidth]{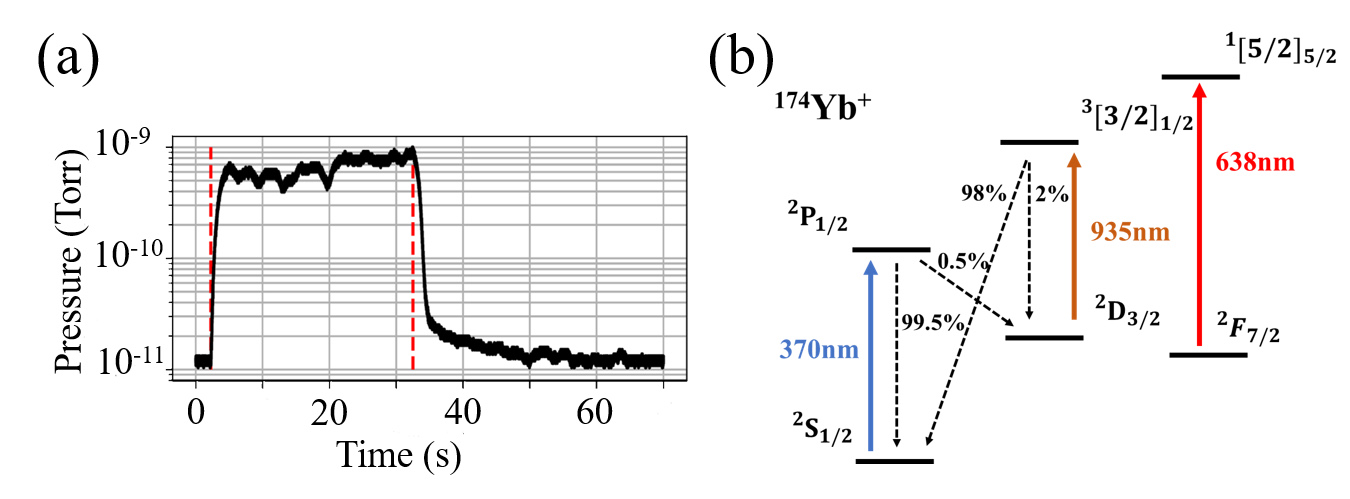}
        \caption[energylevels]{(a) The pressure level of the compact UHV chamber (estimated from the ion pump current and electrode geometry) during a \SI{30}{\second} period of continuously running the ablation laser with a pulse repetition rate of \SI{20}{\hertz} and typical experimental parameters for trapping. The laser is turned on at the first red dotted line and turned off at the second one. (b) Relevant energy levels of $^{174}$Yb$^+$, showing the cooling (\SI{370}{\nano\meter}) and pumping (\SI{638}{\nano\meter}, \SI{935}{\nano\meter}) transitions, and the branching ratio of decay processes.}
        \label{fig:energylevels}
\end{figure}

A schematic diagram of the UHV package is shown in Fig.~\ref{fig:visit}(b). The internal volume of the UHV package contains two small pockets, located approximately \SI{9}{\milli\meter} from the center of the main chamber.  Each of the pockets is loaded with a pellet of 99.9$\%$ purity metallic Yb with a dimension of \SI{1}{\milli\meter} diameter and \SI{2}{\milli\meter} length.  We generate neutral Yb flux by focusing a $\SI{1064}{\nano\meter}$ Q-switched Nd:YAG pulsed laser onto the Yb source pellet\cite{Vrijsen:19} with an incident beam waist of $\SI{180}{\micro\meter}$.  The maximum pulse energy used in this experiment is \SI{0.3}{\milli\joule}, with a temporal pulse length of \SI{8}{\nano\second}. The graph in Fig.~\ref{fig:energylevels}(a) shows the time-dependence of the pressure inside the UHV package during a \SI{30}{\second} ablation loading window as estimated from the ion pump current.  The pressure increase during the ablation window demonstrates that the ablation laser removes material from the Yb pellet, and the ion pump current returns to baseline within seconds after the ablation laser is turned off.

A micro-fabricated surface trap (Sandia National Laboratories HOA~2.0~\cite{Maunz:2016}) is mounted in the center of the main chamber in the UHV package. The trap creates a quasi-static potential by a combination of DC and RF electric fields\cite{Leibfried:2003}.  The neutral atomic flux generated by the ablation laser crosses the trap axis in the center of the trap. Here, the atoms are photoionized by \SI{399}{\nano\meter} and \SI{391}{\nano\meter} laser beams via a two-photon process. Ions with sufficiently low kinetic energy are subsequently trapped by the quasi-static potential created by the voltage on the surface trap's RF and DC electrodes.

The relevant energy diagram of a $^{174}$Yb$^+$ ion is shown in Fig.~\ref{fig:energylevels}(b). The ions are Doppler cooled using the $^2$S$_{1/2} \rightarrow ^2$P$_{1/2}$ transition.  Optical re-pumping from the meta-stable $^2$D$_{3/2}$ state back to the Doppler cooling cycle is done by a \SI{935}{\nano\meter} repump laser. A second repump laser at \SI{638}{\nano\meter} is also applied to remove ions from the meta-stable $^2$F$_{7/2}$ state. The CW lasers all co-propagate in the same photonic crystal fiber and are focused onto the ions by a parabolic mirror. The emitted photons from resonant scattering at \SI{370}{\nano\meter} are collected by an imaging lens with a large numerical aperture ($NA = 0.6$). An electron multiplying CCD (EMCCD) camera (Andor iXon Ultra 888) detects these collected photons to image the trapped ions.

The quality of the vacuum in a trapped ion system dictates the lifetimes of the ion chain. The elastic collision rate between the residual background gas molecules and the trapped ions is a critical parameter of a trapped ion-based quantum computer, since the these collisions with sufficient kinetic energy transfer can significantly disrupt the trapped ion chain. To reliably maintain a chain of ions over the periods of time required for quantum computation, the level of vacuum must be in the UHV regime ($\sim\SI{e-11}{\torr}$)\cite{zhang2017}.

The ion-molecule interaction is modeled by the electric field of the trapped ion polarizing the electron distribution of a background gas molecule to create an induced dipole moment\cite{pagano_cryogenic_2018, hankin_systematic_2019}. This dipole experiences a net attractive force from the ion, and will collide with the ion provided its trajectory passes sufficiently close to the ion.  The ion-molecule interaction potential is
\begin{equation}
    U\left(r\right) = -\frac{C_4}{r^4}
\end{equation}
where $C_4 = \alpha Q^2/8\pi\epsilon_0$, $Q$ is the net charge of the ion, $\alpha$ is the polarizability of the background molecule, and $\epsilon_0$ is the vacuum permittivity.  The collision energy in the center-of-mass frame is given by $E = \mu v^2/2$, where $\mu = m_i m_m/(m_i+m_m)$ is the reduced mass, $m_i$ and $m_m$ are the mass of the ion and the molecule, respectively, and $v$ is the relative velocity between the ion and molecule. The critical impact parameter of ion-molecule collision is then given by $b_c = \left(4C_4/E\right)^{-4}$\cite{Grier2009,Harter2014}.  The Lagrangian equations of motion result in a radial equation governed by the effective potential $U_\text{eff} = U(r) + L^2/2 \mu r^2$ where $L$ is the angular momentum of the system.  When the collision energy is large enough to overcome the barrier between the attractive region (small $r$) and the repulsive centrifugal region (large $r$), a short-range collision occurs.  The Langevin cross-section of this collision process is given by\cite{Langevin1905}
\begin{equation}
    \sigma_L = \pi b_{c}^2 =\pi \sqrt{4C_4/E} =\pi \left(\frac{8}{\mu v^2}\frac{\alpha Q^2}{8\pi \epsilon_0}\right)^{1/2}.
\end{equation}

Considering the situation where the ion collides with background molecules of density $n$, the rate of ion-molecule collision in close range is denoted as $\gamma = n v \sigma_L$, where the density $n$ of the background gas can be expected to obey the ideal gas law $n = P/k_B T$ at the low densities expected under UHV conditions. Here, $k_B$ is the Boltzmann constant, and $P$ and $T$ are the background gas pressure and temperature, respectively. The collision rate is then related to the background pressure by
\begin{equation} \label{eq:collisionrate}
    \gamma = \frac{P Q}{k_B T}\sqrt{\frac{\pi \alpha}{\mu \epsilon_0}}.
\end{equation}

To determine the energy transferred from the molecule to the ion during the elastic collision we calculate the velocity of the ion after the collision as
\begin{equation}
    \vec{v_i} = \frac{m_m \vec{v}}{m_m + m_i}\left(1-\cos\theta \right) + \frac{m_m v \hat{b}}{m_m + m_i}\sin\theta,
\end{equation}
where $\theta$ is the scattering angle and $\hat{b}$ is the unit vector pointing in the direction perpendicular to the initial velocity vector $\vec{v}$. On average, the ion's energy after the collision is 
\begin{equation} \label{eq:collisionenergy}
    \left<E_i\right>_\theta = \left< \frac{m_i v_i^2}{2} \right>_\theta = \frac{2m_im_m}{\left(m_i + m_m\right)^2} E_m
\end{equation}
where the brackets indicate an average over $\theta$, and $E_m = m_m v^2/2$ is the initial kinetic energy of the molecule. We consider the case of a $^{174}$Yb$^+$ colliding with a background H$_2$ molecule, which is the dominant remaining constituent in a well-treated vacuum package, at room temperature. An estimated energy transfer averaged over the molecule’s Boltzmann distribution is $\left<E_i\right>_\theta \approx\SI{0.87}{\milli\electronvolt}$.

\begin{figure}
        \centering
        \includegraphics[width=\columnwidth]{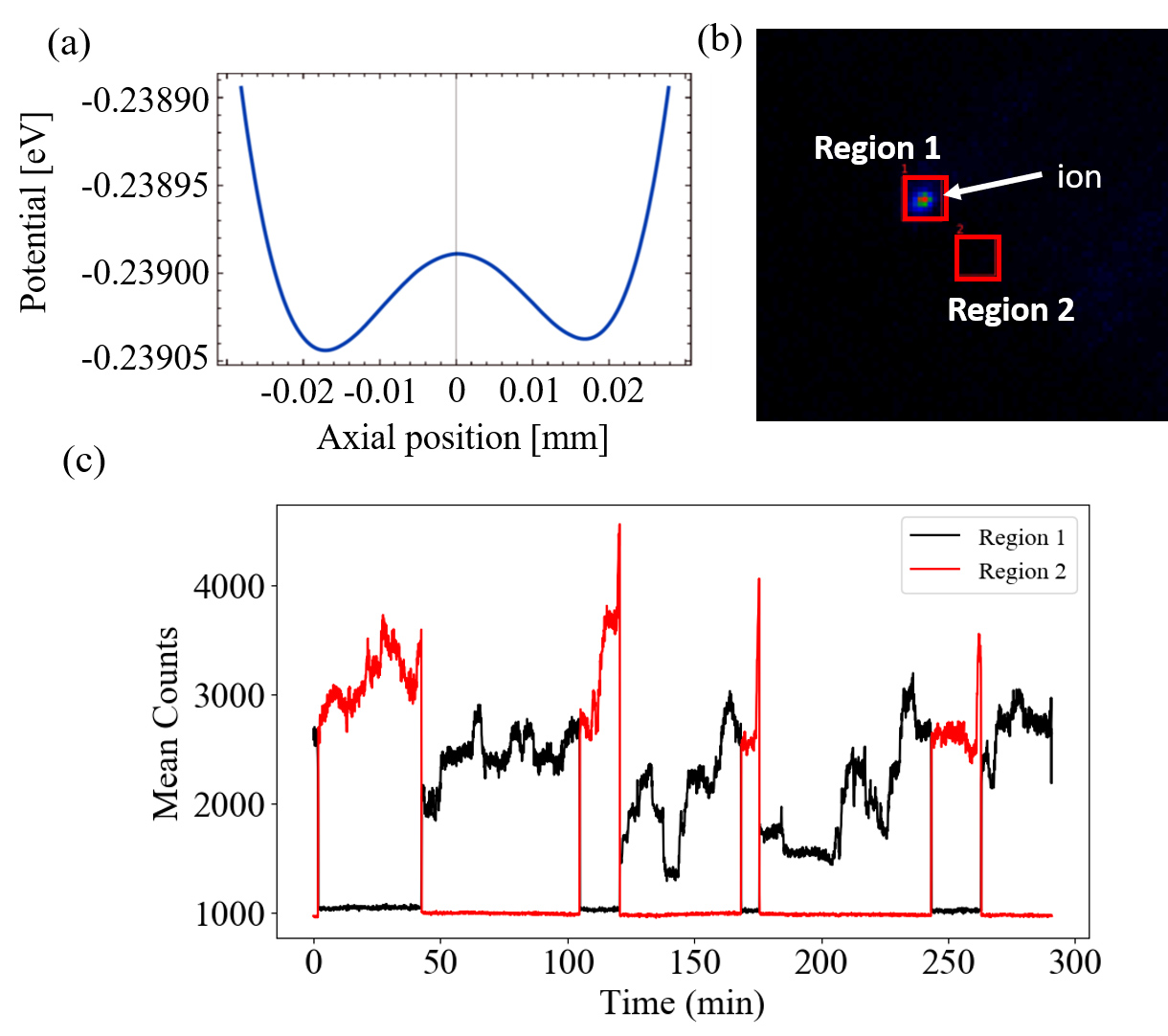}
        \caption[Double-well hopping rate]{(a)~Simulated axial dependence of the double-well potential. The central barrier height is \SI{50}{\micro\electronvolt}.  (b)~Image of the ion, including the regions where the minima of the two potential well are located. (c)~Sample trace of the total EMCCD signal counts for the pixels in regions 1 and 2, showing the position of the ion as a function of time.}
        \label{fig:doublewell}
\end{figure}
In the previous works, the vacuum pressure of trapped ion systems has been estimated by monitoring the rate of elastic collision between ions and background gas leading to reordering of the position of ions. Here, following the method described in the article\cite{pagano_cryogenic_2018, hankin_systematic_2019}, we conduct two experiments to characterize the vacuum level of the compact UHV chamber.  First, we measure the rate at which an ion trapped in a double-well potential hops from one well to the other, and use Eq.~\ref{eq:collisionrate} to relate this hopping rate to the background pressure.  Second, we monitor the rate of reordering events in an ion chain containing Yb ions of two different isotopes. 

The double-well potential is implemented such that the ion can hop between the two potential minima for nearly all collision events with background molecules. To achieve this condition, the height of the potential barrier between the two minima has to be much lower than the average energy transfer due to a collision event. We set the potential barrier height to \SI{50}{\micro\electronvolt}, about 17 times lower than the average collision energy $\left<E_i\right>_\theta$. The potential barrier is designed by the trap simulation code provided by Sandia National Laboratories and the simulated DC potential along the axial axis is shown in Fig.~\ref{fig:doublewell} (a). The potential barrier height was verified by applying a DC voltage to tilt the axial potential until the ion slides from one well to the other. Under these conditions, we expect every collision event to randomize the ion location after the collision.

The hopping rate in the double-well potential is measured by imaging the ion location on the EMCCD camera to determine which of the two wells contains the ion, as shown in Fig.~\ref{fig:doublewell} (b).  The mean time between hopping events, which can be extracted from time traces similar to the one shown in Fig.~\ref{fig:doublewell}(c), is 1 event per 32 $\pm$ 7.9 minutes. The actual collision rate is expected to be double the measured hopping rate, since the ion will eventually get Doppler cooled into either well after the collision, and events where the ion settles back into the original well are not detected using our measurement scheme.  Thus, substituting $\gamma = 1/(\SI{16}{\minute})$, $\mu \approx m_{\rm{H_2}} = \SI{3.32e-32}{\kilo\gram}$, $\alpha_{\rm{H_2}} = \SI{8e-31}{\cubic\meter}$ and $T = \SI{300}{\kelvin}$ into Eq.~(\ref{eq:collisionrate}), we obtain the estimated pressure at the ion location as $P = (2.2 \pm 0.7) \times \SI{e-11}{\torr}$.  This vacuum level is comparable to a well-prepared conventional UHV chamber.

\begin{figure}
        \centering
        \includegraphics[width=\columnwidth]{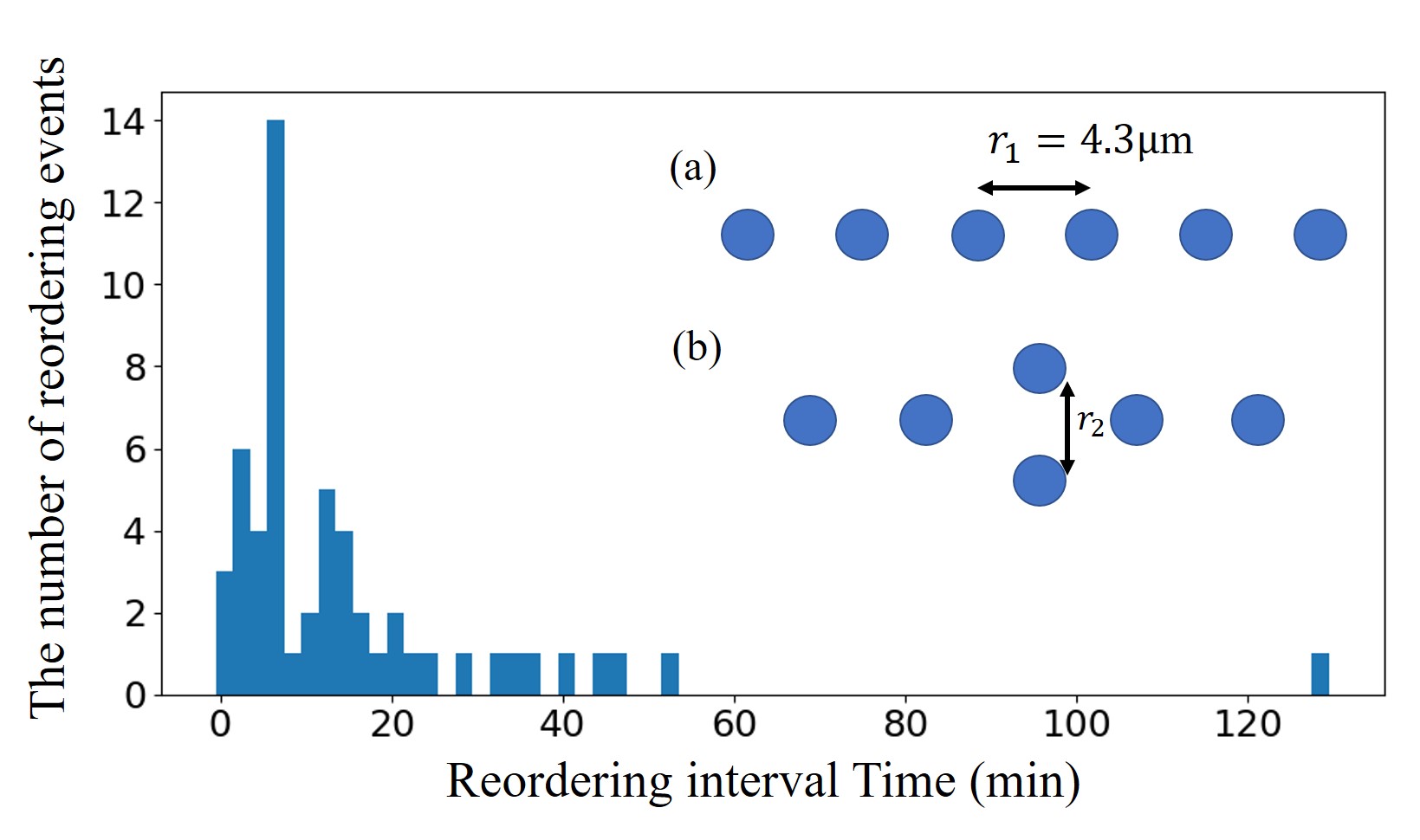}
        \caption[ion hopping]{Histogram of the time interval between ion reordering events. The width of the histogram bin is 2 minutes. 54 reordering events are observed in 15 hours and the average hopping rate is one hopping per 16 min. (inset) The two configurations of the ion chain. To estimate the energy difference between the two configurations, we minimize the energy of the chain as a function of r1 and r2, and we find an energy difference of \SI{1.2}{\milli\electronvolt.}}
        \label{fig:reordering}
\end{figure}

Another method to estimate the pressure is by measuring the rate of reordering events in an ion chain. To estimate the collision energy required to observe a reordering event in a 6-ion chain, we calculate the energy difference between two different chain configurations: one where the ions are aligned along the trap axis in their typical configuration, and the other where two of the ions have been pushed into one of the transverse axis, as shown in the inset of Fig.~\ref{fig:reordering}.
When identical ions are stationary in the trap, the total energy of the chain is given by
\begin{equation}
\begin{split}
    U = \frac{m}{2} \sum^N_{i = 1}\left(\omega_x^2X_i^2+\omega_y^2Y_i^2+\omega_z^2Z_i^2\right) \\
    + \frac{Q^2}{8\pi \epsilon_0}\sum^N_{\substack{i,j=1\\ i \ne j}}\frac{1}{\left|R_i-R_j\right|},
\end{split}
\end{equation}
where $m$ is the mass of the ions, $\omega_{x,y,z}$ are the trap frequencies in the $x$, $y$, and $z$ directions, respectively, $R_i = \left(X_i, Y_i, Z_i\right)$ is the equilibrium position of the $i$-th ion in the trap and $Q$ is the charge of the ions. Using our experimental parameter ($\omega_{x,y} = \SI{2.7}{\mega\hertz}$, $\omega_z = \SI{0.32}{\mega\hertz}$), we obtain an energy difference between these two configuration to be about \SI{1.2}{\milli\electronvolt}.  Therefore, if the ion chain gained an energy in excess of about \SI{1.2}{\milli\electronvolt} from a collision event, a reordering event can occur. Comparing this potential difference to the average energy transferred to an ion in a collision, we expect about 25$\%$ of collisions to transfer sufficient energy to the chain for reordering to occur.

To measure the rate of re-ordering events, we trap a chain of 6 ions containing two different isotopes of Yb: four $^{174}$Yb$^+$ ions and two $^{172}$Yb$^+$ ions. The small difference in mass should not significantly change the expected energy barrier of \SI{1.2}{\milli\electronvolt}.  Due to the isotope shift, only one of the isotopes will be resonant with the \SI{370}{\nano\meter} Doppler cooling laser, and therefore be visible on the EMCCD camera. In this experiment, we choose the four $^{174}$Yb$^+$ as the bright ions while the $^{172}$Yb$^+$ ions remain dark.

A histogram showing the time interval between ion reordering events is shown in Fig.~\ref{fig:reordering}.  We record a video of the ion chain configuration under constant Doppler cooling for a continuous 15 hour period during which we observe 54 reordering events. The width of each time bin is \SI{2}{\minute}. The observed average reordering rate is one event per 16 $\pm$ 2.7 minutes. For 33 out of 54 reordering event only one of the two dark ions moved positions, while for the remaining events both of the dark ions are seen to move. A chain composed of 4 bright and 2 dark ions has 15 distinguishable ion chain configurations.  Therefore, assuming ion positions are completely randomized after a reordering event, we expect a 14/15 (93.3\%) chance to observe the event.  Out of these 14 observable events, 8 of them (57.1\%) will appear to have a single dark ion change position in the chain.  This expected probability matches well with the observed probability of 33/54 (61.1\%).  We can therefore effective consider the chain order randomized every time a chain reordering event occurs.

We expect the collision rate for 6 ions to be 6 times the rate for a single ion, and therefore we expect a collision rate of $1/(2.7 \pm 0.7\SI{}{\minute})$ for the 6-ion chain based on the result of the double-well experiment. Considering the potential energy difference between the two configurations shown in Fig.~\ref{fig:reordering} is \SI{1.2}{\milli\electronvolt} and 14 out of 15 reordering events are observable, the true rate of collision events is estimated as $1/(3.7 \pm 1.0\SI{}{\minute})$ where the error bars are due to the combination of the statistical fluctuations in the time between collisions and the uncertainty of the trap RF voltage. The measured collision rates from the two experiments are therefore consistent.

We developed a compact room-temperature trapped ion system, where the chamber volume, as well as the overall footprint of the UHV package, is significantly smaller than the conventional chamber used in similar experiments.  We confirm that the vacuum level of this system is in the \SI{1e-11}{\torr} range which is sufficiently low for most trapped ion experiments.  The compactness of this system provides many opportunities for overall system improvements, including improved temperature stability and mechanical robustness. This compact system has significant potential to enhance the performance of future trapped ion quantum computers, where the fidelity of quantum logic operation is limited by these classical instabilities.     

\begin{acknowledgments}
The authors would like to express our appreciation to Farhad Majdeteimouri and Tianyi Chen for contributions to this project. 
This work is supported by Intelligence Advanced Research Projects Activity through Army Research Office (W911NF16-1-0082). Y. A. is supported by the National Science
Foundation (EFMA-1741651).
\end{acknowledgments}

\section*{Data Availability}
The data that support the findings of this study are available
within this article.

\nocite{*}

\bibliography{VISTref.bib}

\end{document}